\documentclass[pre,twocolumn,showpacs,preprintnumbers,amsmath,amssymb]{revtex4}

\usepackage{graphicx} 
\usepackage{dcolumn}
\usepackage{bm}
\usepackage{psfrag}
\usepackage{epsfig}

\begin{document}

\title{Boltzmann equation approach to transport in finite modular quantum systems}

\author{Mehmet Kadiro\=glu}
\email{mkadirog@uos.de}
\author{Jochen Gemmer}
\affiliation{Department of Physics, University of Osnabr\"uck,
D-49069 Osnabr\"uck, Germany}

\date{\today}

\begin{abstract}
We investigate the transport behavior of finite modular quantum
systems. Such systems
have recently been analyzed by different methods. These approaches indicate diffusive behavior even
and especially for finite systems. Inspired by these results we analyze analytically and
numerically if and in which sense the dynamics of those systems are in
agreement with an appropriate Boltzmann equation. We find that the 
transport behavior of a certain
type of finite modular quantum systems may indeed be described in terms
of a Boltzmann equation. However, the applicability of the Boltzmann
equation appears to be rather limited to a very specific type of model.

\end{abstract}

\maketitle


\section{Introduction}
\label{intro}
There are essentially two major tools which are used to analyze the  transport behavior of quantum systems: Linear response theory as
implemented in terms of the Kubo formula \cite{Kubo,mahan,Klu,Zotos,Heidr,Jung}, and approaches based on the 
Boltzmann equation (BE) \cite{Boltzmann}. The validity of the latter has been subject
to ongoing discussions during the last century \cite{boltzmann2}. This refers to the BE
as a method to describe gas-dynamics on purely classical grounds. In the
context of quantum mechanics the situation may even be more
complicated. Can the dynamics of systems that are controlled by
the Schr\"odinger equation (SE) be mapped on a BE? 
And if so, how? Considerable work in that direction has been done by
Peierls \cite{Peierls},
Kadanoff, Baym \cite{baym} and others \cite{keldysh,winkel,Kohn}. In this literature it is frequently pointed out that the
mapping typically relies on additional assumptions like the ``random phase
approximation'' \cite{Peierls} or the possibility to truncate graphic expansions that
does not necessarily follow from the underlying dynamics \cite{baym}. 
But also recent publications address the mapping of quantum dynamics
onto BE's \cite{spohn,Horny,vacchi}. 
In the article at hand we investigate the applicability of  a BE to systems which are
both, complex enough to exhibit diffusive behavior and simple enough to
be analyzed from first principles by direct numerical integration. Thus
the dynamics as resulting from a BE may simply be compared to the
dynamics as resulting from the SE. 
The results for transport behavior of those systems obtained by other
methods (Kubo formalism, Hilbert space average method
(HAM), time-convolutionless (TCL) \cite{Buch,TCL,JG}, which have been mentioned in the abstract, may be found in \cite{heat,kubop,chaos,Breu} 

The article at hand is organized as follows:\\
In Sect. \ref{sec-1} we very briefly review the concepts underlaying the
famous BE which is meant to be a gross description
of the  dynamics of dilute gases. We comment on linear forms of the
BE and their diffusive solutions, i.e., we state an explicit form for
the diffusion coefficient.
In Sect. \ref{sec-2} we introduce our finite modular quantum system 
which may be viewed as model for a particle hopping between a few lattice sites
or a model for the energydynamics within a chain of, e.g.,
molecules. The main intention of the work at hand is to investigate
the transport behavior of finite
quantum systems by means of an analysis based on the BE. To those ends 
we propose to identify the classical particle densities which appear in the BE with
the the quantum mechanical occupation numbers of current
eigenstates (\ref{sec-3}). To justify this concept we numerically and analytically
analyze in Sect. \ref{sec-4} whether the dynamics of the current
occupation numbers are indeed well described by an appropriate linear
BE. This turns out to be the case but only for a quite specific form
of the interactions within the model. For this type on interactions we
concretely compute the adequate linear BE and determine the diffusion
coefficient using the form given in Sect. \ref{sec-1}. This coefficient turns out to be
in accord with the results from the different approaches to similar
systems mentioned above \cite{heat,kubop,chaos}.
An ab initio numerical analysis of the full dynamics of the quantum
system (including lattice site occupation numbers)
shows that it indeed exhibits diffusive behavior controlled by the
above diffusion coefficient. In the last Sect. we discuss the
dependence of those results on special properties of our model.

\section{Boltzmann equation and diffusive solutions}
\label{sec-1}

As wellknown, in 1872 Boltzmann undertook to explain the macroscopic dynamics of dilute gases. 
For the description of a gas Boltzmann introduced the
${\bf\mu}$-space, which is essentially a one-particle phase space. An
$N$-particle gas would thus technically be represented by $N$ points in
${\bf\mu}$-space rather than one point in standard Hamiltonian phase
space. But instead of using $N$ points in ${\bf\mu}$-space for the
description of the gas, Boltzmann introduced a distribution function
$\Phi({\bf r},{\bf v},t)$  in a somewhat ''coarse-grained'' ${\bf\mu}$-space 
which is supposed to give the number of particles being in a {\bf $µ$}-space cell around 
$d^{3}rd^{3}v$:=$dxdydzdv_{x}dv_{y}dv_{z}$. Instead of trying to
describe the
motion of every single particle (which is impossible due to 
the huge numbers of particles in a gas) Boltzmann suggested his famous equation which describes the time evolution of $\Phi({\bf r},{\bf v},t)$
in $\mu$-space and is, in the absence of any external force, given by:
\begin{eqnarray}
\frac{d}{dt}\Phi({\bf r},{\bf v})={\bf v}\cdot\nabla_{{\bf r}}\Phi({\bf r},{\bf v})+\dot{\Phi}({\bf r},{\bf v})_{scatt},
\end{eqnarray}   
The first expression on the right-hand-side is supposed to account for
the dynamics due to particles  that do not collide, whereas
$\dot{\Phi}({\bf r},{\bf v})_{scatt}$ describes the dynamics arising
from collisions. Those dynamics are only taken into account in terms
of the transition rates $R$ and the (coarse grained) particle densities
$\Phi$, neglecting all correlations and structures on a finer
scale. Thus this type of dynamics implement the famous 
``assumption of molecular chaos'' or the
so-called ``Sto\ss{}zahlansatz''. (According to the Sto\ss{}zahlansatz
particles are not correlated before collisions, even though they
get correlated by collisions, due to very many intermediate collisions
before the same particles collide again). However, this treatment of
collisions introduces the irreversibility into the BE that is not
present in the underlying Hamiltonian equations.\\
\noindent If either $\Phi$ is close to equilibrium or for systems in which
particels only collide with external scattering centers (no
particle-particle collisions), the BE takes on the following linear form.
\begin{eqnarray}
\dot{\Phi}({\bf r},{\bf v})+{\bf v}\cdot\nabla_{{\bf r}}\Phi({\bf r},{\bf v})=\int R({\bf v},{\bf v'})\Phi({\bf r},{\bf v'})d{\bf v}',
\end{eqnarray}
A ``velocity discretized'' version of this linear BE reads
\begin{eqnarray}
\dot{\Phi}_{i}({\bf r})+{\bf v}_{i}\cdot\nabla_{{\bf r}}\Phi_{i}({\bf r})=\sum_{j}R_{ij}\Phi_{j}(\bf r),
\label{BEdisc}
\end{eqnarray}
\noindent where the matrix of rates $R_{ij}$ is of the standard form as
appearing in master equations. (It is this discrete linear BE which we
are eventually going to use to describe our quantum system.)\\ 
As wellknown, in the limit of small density gradients (``long wavelength
hydrodynamic modes'') (\ref{BEdisc}) may feature diffusive solutions, i.e., solutions that fulfill
\begin{eqnarray}
\dot{\rho}({\bf r})=\kappa\Delta\rho({\bf r}).
\label{diffeq}
\end{eqnarray} 
where the particle density $\rho({\bf r})$ is given by 
\begin{eqnarray}
\rho({\bf r})=\sum\limits_{i}\Phi_{i}({\bf r})
\end{eqnarray}
The diffusion coefficient in (\ref{diffeq}) takes on the form
\begin{eqnarray}
\kappa=-\sum_{i,j}v_{i}R^{-1}_{ij}v_{j}\Phi^{0}_{j}.
\label{DiffCoeff}
\end{eqnarray}
 $\Phi^{0}_{j}$ is the equilibrium velocity distribution (i.e., the
 solution of $R_{ij}\Phi^{0}_{j}=0$, accordingly called
 ``null-space''). $R^{-1}_{ij}$ denotes the inverse of the  rate
 matrix but without the null-space \cite{Balescu,Brenig}. (In the absence of any external
 scatterers $\kappa$ may diverge, indicating that no diffusive bahavior
 can be expected.)
\section{Definition of the model and its diffusive behavior}

\label{sec-2}
The systems we investigate in the following are called ``finite
modular quantum systems''. (This type of system has also been
investigated in\cite{heat,kubop,chaos}). Those systems are essentially meant to allow for an
investigation of transport behavior from first principles rather than
to model some material or real physical system in great detail. 
Their total Hamiltonians may be written as
\begin{eqnarray}
\label{E-2-1}
\hat{H}=\sum_{\mu =1}^{N}\hat{h}(\mu)+\sum_{\mu =1}^{N}\hat{V}(\mu,\mu+1), \qquad\text{with}\quad N+1\widehat{=}1\nonumber\\
\end{eqnarray}
where $\hat{h}(\mu)$ denotes the local Hamiltonian of a subunit $\mu$, $\hat{V}(\mu,\mu+1)$
a next-neighbor interaction between subunits of the ring and $N$ the total number of
subunits in the system.  
\begin{figure}
\centering
\includegraphics[width=75mm]{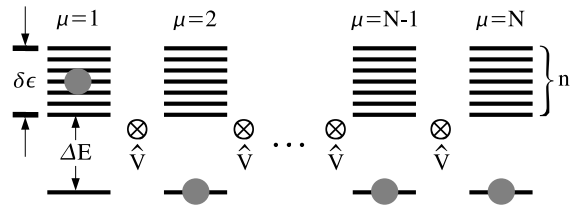}
\vspace*{1mm}
\caption{\label{chain}Sketch of the systems which are investigated for
  transport: N identical weakly
coupled subsystems with a non- degenerate ground state and a band of
equidistant energy levels.}
\end{figure}
A model which can be described by such a Hamiltonian is illustrated in Fig. \ref{chain}.
It consists of $N$ identical subsystems where each subunit features a nondegenerate ground state, a wide
energy gap ($\Delta E$) and  an energy band ($\delta\epsilon$) which 
contains $n$ equidistant energy levels. The local Hamiltonian is defined by
\begin{eqnarray}
\label{loc}
\hat{h}(\mu)=\sum\limits_{i=0}^{n+1}E_{i}\hat{\sigma}^{\dagger}_{i}(\mu)\hat{\sigma}_{i}(\mu) &\quad\text{with}& 
E_{i}=E_{0}+\Delta E+i\frac{\delta\epsilon}{n},\nonumber\\
 &\quad\text{for}& i\neq 0
\end{eqnarray}
and $\hat{{\sigma}}_{i}(\mu)=|0,\mu\rangle \langle i,\mu|$. Here
$|0(i),\mu\rangle$ denotes the ground state (i'th level of the
excitation band) of the $\mu$'th subunit. The interaction is specified
by
\begin{eqnarray}
\label{int}
\hat{V}(\mu,\mu+1)=\lambda\sum_{i,j=1}^{n}\tilde{V}_{ij}\hat{\sigma}^{\dagger}_{i}(\mu)\otimes\hat{\sigma}_{j}(\mu +1)+\text{h.c.}.
\end{eqnarray}  
(h.c. denotes
the hermitian conjugate of the previous sum). The $\tilde{V}_{ij}$ are
complex numbers which are assumed to be adequately normalized such
that $\lambda$ controlls the overall interaction strength. The choice
of these $\tilde{V}_{ij}$ concretely determines the model. In
\cite{heat,kubop,chaos} the $\tilde{V}_{ij}$ are simply chosen to be independent random
numbers (with mean zero) which leads to diffusive dynamics (in a sense
explained below). In the paper at hand it turns out that this choice does
not yield dynamics which are in accord with a BE. In fact, we will
explain in Sect. \ref{sec-4} that it is only a rather specific
choice which produces this accordance. However, at this point we do not specify the
$\tilde{V}_{ij}$ any further.

Irrespective of the choice for the $\tilde{V}_{ij}$ the  system may be viewed as a very simplified model for, e.g., a
chain of coupled molecules or quantum dots, etc. In this case the
hopping of the excitation from one subunit to another corresponds to
energy transport. It may
as well be viewed as a tight-binding model for particles on a
lattice (in second quantization). There are, however, many ($n$)
bands (orbitals per site) but no particle-particle
interaction in the sense of the Hubbard-model (cf. (\ref{E-2-1}), (\ref{int})). Thus one may  characterize this model as a
``single-particle multi-band quantum wire'' with possibly random (interband)
hoppings. Nevertheless, due to the independence of $\tilde{V}_{ij}$, $E_{i}$ of
$\mu$ these are systems without disorder in the sense of, say,
Anderson \cite{Anderson}. (Note that the ``total particle number''
$\hat{N}:=\sum_{i,\mu}^{n,N}\hat{\sigma}^{\dagger}_{i}(\mu)\hat{\sigma}_{i}(\mu)$
is conserved)\\ 
\noindent We define diffusive transport for this model based on the
evolution of expectation values
$q_{\mu}(t)=\langle\Psi(t)|\hat{q}(\mu)|\Psi(t)\rangle$ of
local fractions of some (globally conserved) quantity like energy or
particle number ($|\Psi(t)\rangle$ denotes the systems full state). I.
e., one may consider energy transport:
$\hat{q}(\mu)=\hat{h}(\mu)$ or particle transport:
$\hat{q}(\mu)=\hat{n}(\mu)=\sum_{i=1}^{n}\hat{\sigma}_{i}^{\dagger}(\mu)\hat{\sigma}_{i}(\mu)$.,
etc. We will call the transport diffusive if those local expectation values obey
\begin{eqnarray}
\dot{q}_{\mu}(t)=\kappa\cdot\left(q_{\mu+1}(t)+q_{\mu-1}(t)-2q_{\mu}(t)\right),
\label{rateequation}
\end{eqnarray}
which is a discrete version of the diffusion equation
(\ref{diffeq}). In this case obviously only the diffusion coefficient
$\kappa$ remains to be computed. If ``diffusivity'' is supposed to be a property of a
system (\ref{rateequation}) has to apply of course to the largest part
of all initial states under consideration. 
\section{Choice of quasiparticles for modelling transport}
\label{sec-3}

It is not a priory clear which quantum observable could play the role
of a classical particle density in phase space when one attempts to map the
quantum dynamics of some system onto a BE. A minimum requirement is surely the clear
correspondence to some velocity\\
\noindent In the context of heat transport through electrically isolating
crystals one identifies the particle (phonon) density with the average
occupation number of a phonon mode. The velocity is then extracted
from the phonon dispersion relation \cite{baym,Peierls}. However, the phonons are eigenmodes 
of a harmonic chain (spring-and-ball-model), and the scattering
arises from a (weak) anharmonic part of the interaction. In our model
the interactions do not allow for a decomposition into an harmonic and
an anharmonic part. Thus this scheme cannot be applied here.\\
\noindent In the context of particle transport through systems of interacting
particles in periodic lattices, the (quasi)particle density is basically
identified with the occupation number of eigenmodes (Bloch waves) of the
interaction-free particle, e.g., the crystal electron. The velocity
(group velocity) is
then computed from the corresponding dispersion relation . Since our
model (from a ``Hubbard'' point of view) features no
particle-particle interaction  such a description would result in a BE
without any scattering at all ($R_{ij}=0$), hence in ballistic transport. This
however is not in accord with our findings (see below).\\
\noindent Thus, here we suggest to identify the particle density with the occupation
number of eigenmodes of a suitable current operator. To define a current operator 
on the basis of the transported quantity we consider the time evolution of the 
corresponding local operator $\hat{q}(\mu)$ at site $\mu$ which is given by the Heisenberg
equation of motion \cite{Klu,Heidr,Jung,Michi}:
\begin{eqnarray}
\frac{d}{dt}\hat{q}(\mu)&=&\frac{\partial}{\partial t}\hat{q}(\mu)+\frac{{\bf i}}{h}\left[\hat{H},\hat{q}(\mu)\right]= \frac{{\bf 
i}}{h}\left[\hat{H},\hat{q}(\mu)\right],\nonumber\\          
\label{heimo}
\end{eqnarray}
since the operators $\hat{q}(\mu)$ mentioned above are explicitly time independent. After inserting equation (\ref{E-2-1}) and 
applying the explicit form of $\hat{q}(\mu)(=\hat{h}(\mu),\hat{n}(\mu))$ we obtain:
\begin{eqnarray}
\frac{d}{dt}\hat{q}(\mu)=\frac{{\bf i}}{h}\left[\hat{V}(\mu-1,\mu),\hat{q}(\mu)\right]+\frac{{\bf 
i}}{h}\left[\hat{V}(\mu,\mu+1),\hat{q}(\mu)\right].\nonumber\\
\label{heimo2}
\end{eqnarray}
If conserved quantities $q(\mu)$ are considered, currents are routinely defined on the basis of the temporal change
of the respective densities by means of a (discrete) continuity equation which reads for $\hat{q}(\mu)$:
\begin{eqnarray}
\frac{d}{dt}\hat{q}(\mu)=\hat{j}(\mu,\mu +1)-\hat{j}(\mu-1,\mu)=-\text{div}\hat{j}.
\label{div}
\end{eqnarray}
Comparing Eq.(\ref{heimo2}) with Eq.(\ref{div}) this suggests the definition of a local current operator
\begin{eqnarray}
\hat{j}^{Q}(\mu,\mu +1)=\frac{{\bf i}}{h}\left[\hat{V}(\mu,\mu+1),\hat{q}(\mu)\right],
\end{eqnarray} 
whereas the total currentoperator $\hat{J}^{Q}$ is given by
\begin{eqnarray}
\hat{J}^{Q}=\sum\limits_{\mu =1}^{N-1}\hat{j}^{Q}(\mu,\mu +1)=\frac{{\bf i}}{\hbar}\sum\limits_{\mu =1}^{N-1}\left[\hat{V}(\mu,\mu 
+1),\hat{q}(\mu)\right]
\end{eqnarray}
E.g., in the case of energy transport the current 
operator is 
\begin{eqnarray}
\hat{J}^{H}=\frac{{\bf i}}{\hbar}\cdot\lambda\sum\limits_{\mu 
=1}^{N-1}\sum\limits_{i,j}^{n}E_{i}\tilde{V}_{ij}\hat{\sigma}_{i}^{\dagger}(\mu)\otimes\hat{\sigma}_{j}(\mu+1)+\text{h.c.}.
\label{energycurrent}
\end{eqnarray}
Since the current is a product of velocity and density we assign a
velocity to our 
quasiparticles by the relation
\begin{eqnarray}
v_n=\frac{j^{Q}_{n}}{\langle j^{Q}_{n}|\hat Q|j^{Q}_{n}\rangle} \quad\text{with}\quad Q=\sum\limits_{\mu =1}^{N}\hat{q}(\mu),
\label{velo}
\end{eqnarray}
where $j^{Q}_{n}$ is an eigenvalue of the corresponding current operator.
In this way the velocities that eventually appear in (\ref{DiffCoeff}) may be defined.

\section{Analysis of the dynamics of the model}   
\label{sec-4}

\begin{figure}
\centering
\includegraphics[width=80mm]{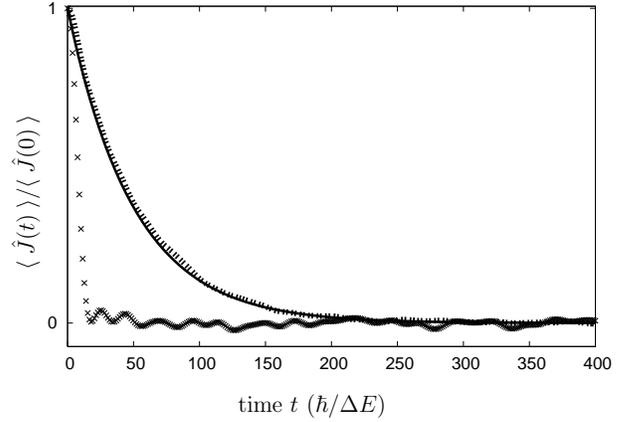}
\vspace*{1mm}
\caption{\label{decay}Expectationvalue of the currentoperator obtained from the exact solution of the
SE (here $|\psi(0)\rangle$ is a random pure state for which only the 2625 current eigenstates with the largest eigenvalues feature non-zero
amplitudes) (dashed line). This is compared with the exponential function (solid line) featuring
the relaxation time $\tau_{R}$ one obtains from an analytical analysis of $\tilde{\mathcal{C}}(t)$ (cf. (\ref{auto})).
System parameters: N = 3, n = 3500, $\Delta E = 1.0$,
$\delta\epsilon$ = 0.35, $\lambda = 5\cdot 10^{-5}$, $\alpha = 5\cdot10^{-3}$.
The pointed curve shows the current expectationvalue of a model
without the proposed weighting function (cf. (\ref{Cauchy})) the other
system parameters 
are the same as mentioned above. The current relaxes, but not in an
exponential way (see also Fig. \ref{Logdecay}).}
\end{figure}

\begin{figure}
\centering
\includegraphics[width=80mm]{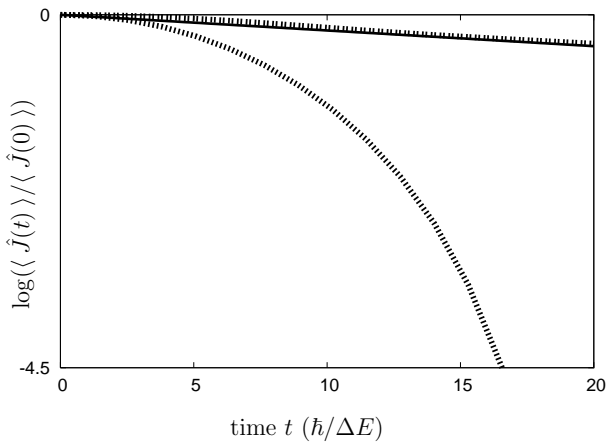}
\vspace*{1mm}
\caption{\label{Logdecay}
Logarithmic plot of the current expectationvalues shown in
Fig. \ref{decay}. The straight dashed line corresponds to the
numerical result for the modified interaction, the parabolic dashed
line to the numerical result for a purely random interaction (the
solid line corresponds to theory). The purely random interaction does
not yield any exponential decay of the current.}
\end{figure}

In this Sect., we analyze whether the above described choice of
quasiparticles is in accord with a BE from a dynamical
point of view. Or, to formulate concise questions: May the dynamics of the
populations of the current eigenmodes as resulting from the
Schr\"odinger-dynamics of the quantum model be described in terms of an
adequate BE? And if so what would be the rates in the scattering term?
\\ 
In Sect. \ref{sec-3} we suggested to identify the particle
density as appearing in the BE by a quantity that corresponds to a specific
velocity but not to a spatial
coordinate. This is obviously in some sense insufficient since
particle densities in phase space are labeled by velocity and
poisition. However, since the model features translational invariance, the
current eigenmodes (the populations of which are supposed to
correspond to the particle densities) stretch uniformly over the full
model. Thus the dynamics of their populations may be expected to possibly
correspond to the dynamic as resulting from a BE for particle
densities that are uniform with respect to the position coordinate, i.e., $\nabla_{\bf r}\Phi =0$. In this case particle densities are only labled
by velocities and may directly be identified with current eigenstate
populations. In the above mentioned case, (\ref{BEdisc}) simplifies to  
\begin{eqnarray}
\dot{\Phi}_{n}(t)=\sum_{m}R_{nm}\Phi_{m}(t).
\label{BEdischom}
\end{eqnarray}
The above equation (\ref{BEdischom}) yields exponential decay for the
$\Phi$'s (possibly with various relaxation times). In the following we
investigate whether the same behavior results from the SE
for the current eigenmode populations i.e., if we
compute $\Phi_n(t)$ from the definition
$\Phi_n$:=Tr$\left\{\hat{\rho}(t)\hat{P}_{n}\right\}$ where $\hat{P}_{n}=|j_{n}\rangle\langle j_{n}|$ is the projector onto
the subspace spanned by the current eigenstate $|j_{n}\rangle$. If
this is the case the SE and the BE may be in
accord. 
However, this
question is hard to answer in general without using numerical
reasoning, since otherwise the quantum dynamics for
$\Phi_n(t)$ cannot be found. Thus, rather than analyzing the
dynamics of the $\Phi_n(t)$'s themselves, we analyze a
function of those that can be estimated without using
numerics. Strictly speaking this of course means we go from a proof to
a check of consistency (However, we also check dynamics of the
$\Phi_n$ also directly numerically). This function we call $\mathcal{C}(t)$ and construct it as
\begin{eqnarray}
\mathcal{C}(t)=\sum_{n}j_{n}\cdot \text{Tr}\left\{\hat{\rho}(t)\hat{P}_{n}\right\}=\text{Tr}\left\{\hat{J}\hat{\rho}(t)\right\},
\label{TraceJJ}
\end{eqnarray}
It is obviously just a weighted sum of the $\Phi_{n}(t)$'s. If
its dynamics are in accord with (\ref{BEdischom}), $\mathcal{C}(t)$
should also decay  exponentially. As (\ref{TraceJJ}) shows,
$\mathcal{C}(t)$ is simply the current expectationvalue. In the
Heisenberg picture the latter reads
$\mathcal{C}(t)=$Tr$\left\{\hat{J}(t)\hat{\rho}(0)\right\}$. Again, we cannot
analyze this in full generality, thus we specialize to a concrete
initial state, which is sometimes called a ``deviation density
matrix''. It is given by 
$\hat{\rho}(0)=d^{-1}\cdot\hat{{\bf 1}}+\epsilon\hat{J}(0)$ 
($\hat{{\bf 1}}$ = identity, $d$ = dimension of the corresponding (sub-)space) and fulfills the relation 
Tr$\left\{\hat{\rho}_{0}\right\}=1$ for density operators due to the fact that the current operator 
is traceless.  Thus we obtain $\mathcal{C}(t)=\epsilon\cdot$Tr$\left\{\hat{J}(t)\hat{J}(0)\right\}$, 
which is simply the current-autocorrelation function. Without going into any detail
here we should mention that, following concepts based on the Hilbert
space average method (HAM) as presented in \cite{kubop,JG,Breu}, $\mathcal{C}(t)$ can be
expected to reasonably describe the evolution of the expectation value
of the current for almost any initial state. Thus the results which
will be derived analytically below can safely be expected to apply to a much larger
class of initial states than covered by the deviation density
matrix. Especially the results can be expected to apply to the largest
part of all pure states, which is the class of states which will be
primarily analyzed numerically below.\\
\noindent The current-autocorrelation function reads:
\begin{eqnarray}
\tilde{\mathcal{C}}(t)= \sum\limits_{\alpha,\beta=1}^{n}|\langle \alpha|\hat{J}(0)|\beta\rangle|^{2}e^{-\frac{{\mathbf 
i}}{\hbar}(E_{\alpha}-E_{\beta})\cdot t},
\label{auto}
\end{eqnarray}
where $|\alpha(\beta)\rangle,E_{\alpha(\beta)}$ are energy eigenvectors respectively
eigenvalues of the full, coupled system.\\
\noindent Here and in the following we restrict ourselves to the ``one
excitation'' (one-particle) subspace. This is possible since the
particle number is conserved, cf. (\ref{int}). If the coupling ($\lambda$)
is weak, it may be reasonable to approximate the true
eigenvectors/eigenvalues of full system
$|\alpha (\beta)\rangle,E_{\alpha(\beta)}$ that appear explicitly in the correlation
function, by the eigenvectors/eigenvalues of the uncoupled system from
the one-particle subspace which feature the particle at a given site $\mu$. Since the current operator 
only ``couples'' states featuring the particle in adjacent
sites, the double sum over sites collapses and we find in this approximation 
\begin{eqnarray}
\mathcal{\tilde{C}} (t)&\approx& \sum\limits_{\mu=1}^{N}\sum\limits_{i,j=1}^{n}|\langle 
i,\mu|\hat{J}(0)|\mu+1,j\rangle|^{2}e^{-\frac{{\mathbf 
i}}{\hbar}(E_{i}-E_{j})\cdot t}.\nonumber\\
\end{eqnarray}
For $\hat{J}(0)$ we plug in the energycurrent operator as given by
(\ref{energycurrent}). Obviously the addends do not depend on $\mu$, thus performing
the corresponding sum simply results in a prefactor $N$. If we assume
$\Delta E>>\delta\epsilon$ and thus $E_i\approx \Delta E$ for the 
current operator (not for the exponential)we get
\begin{eqnarray}
\tilde{\mathcal{C}}(t)=\gamma \sum\limits_{i,j=1}^{n}|\tilde{V}_{ij}|^{2}e^{-\frac{{\mathbf
i}}{\hbar}\frac{\delta\epsilon}{n}(i-j)\cdot t} \nonumber \\
\end{eqnarray}  
where $\gamma:=2N\left(\frac{\lambda}{\hbar}\Delta E\right)^{2}$(For
particle transport we simply have to set $\Delta E=1$).
 In order to evaluate this expression we split up the double sum into two
double sums. In the first double sum we perform the index
transformation $i=k-l+1, j=n-l+1$, in the second the index
transformation $i=n-l+1, j=k-l+1$. Thus using for $\tilde{V}_{ij}$ the
definition from (\ref{Cauchy}) yields:
\begin{eqnarray}
\tilde{\mathcal{C}}(t)=&\gamma&\cdot\left\{\sum\limits_{k=1}^{n}A(k,n)e^{-\frac{{\mathbf i}}{\hbar}\frac{\delta\epsilon}{n}(k-n)\cdot 
t}\right.\nonumber\\ 
 &+&\left. \sum\limits_{k=1}^{n-1}A'(k,n)e^{\frac{{\mathbf i}}{\hbar}\frac{\delta\epsilon}{n}(k-n)\cdot t}\right\},
\end{eqnarray}
\begin{eqnarray}
\text{with}\quad &A(k,n)&=\sum\limits_{l=1}^{k}|\tilde{V}_{k-l+1,n-l+1}|^{2}\nonumber\\ 
\quad\text{and}\quad &A'(k,n)&=\sum\limits_{l=1}^{k}|\tilde{V}_{n-l+1,k-l+1}|^{2}.
\end{eqnarray}
Hence $\tilde{\mathcal{C}}(t)$ is essentially the Fourier transform of
the $A,$ $A'$. As explained above, if the quantum dynamics are claimed to
be in accord with the BE, $\tilde{\mathcal{C}}(t)$ must decay
exponentially. But this  will only be the case if  $A$, $A'$ take the
form of some Lorentzian in the argument $(k-n)$. If, however, the
$\tilde{V}_{ij}$ are chosen to be independent (gaussian) random numbers
as done in \cite{heat}, the $A$, $A'$ will simply be proportional to $k$ and
thus no exponential decay of $\tilde{\mathcal{C}}(t)$ is predicted
within the framework of this approach. This expectation is confirmed
by the numerical computation of  $\tilde{\mathcal{C}}(t)$ as resulting
from the Schr\"odinger equation (see Fig.'s \ref{decay}, \ref{Logdecay}). Thus, in general, the
dynamics of the quasiparticles (at least for this definition of
quasiparticles) cannot be claimed to be in accord with a BE. If one
modifies the weights of the $\tilde{V}_{ij}$, however, one can enforce
a Lorentzian shape upon $A$, $A'$. We now choose 
$\tilde{V}_{ij}$ as 
\begin{eqnarray}
\tilde{V}_{ij}=\frac{V_{ij}}{\sqrt{(1-\frac{1}{n}|i-j|)\cdot(1+\alpha^2(i-j)^2)}},
\label{Cauchy}
\end{eqnarray}
where $V_{ij}$ are still randomly distributed complex numbers
normalized to $n^{-2}\sum_{i,j}|V_{ij}|^{2}=1$ and $\alpha$ is an (to
some extend) arbitrary parameter. This choice yields in good
approximation for large $n$: $A\approx A'\approx n\cdot(1-\alpha^{2}(k-n)^2)^{-1}$
and thus leads to
\begin{eqnarray}
\tilde{\mathcal{C}}(t)\approx n\gamma\cdot\left\{\sum\limits_{k=1}^{n}
\frac{e^{-\frac{{\mathbf i}}{\hbar}\frac{\delta\epsilon}{n}(k-n)\cdot t}}{1+\alpha^{2}{(k-n)}^2}
+\sum\limits_{k=1}^{n-1}\frac{e^{\frac{{\mathbf i}}{\hbar}\frac{\delta\epsilon}{n}(k-n)\cdot t}}{1+\alpha^{2}{(k-n)}^2}\right\}. \nonumber \\
\end{eqnarray}        
If $\alpha << 1$ we may replace the sums by the following integrals
\begin{eqnarray}
&\tilde{\mathcal{C}}(t)&\approx \nonumber\\ 
&n\gamma&\cdot\left\{\int\limits_{1}^{n}\frac{e^{-\frac{{\mathbf i}}{\hbar}\frac{\delta\epsilon}{n}(k-n)\cdot 
t}}{1+\alpha^{2}{(k-n)}^2}dk+\int\limits_{1}^{n-1}\frac{e^{\frac{{\mathbf i}}{\hbar}\frac{\delta\epsilon}{n}(k-n)\cdot
t}}{1+\alpha^{2}{(k-n)}^2}dk\right\}, \nonumber\\
\label{cossum}
\end{eqnarray}
If furthermore $(\alpha n)^2 >>1$ we may take the lower  limit to
negative infinity, obtaining
\begin{eqnarray}
\label{expz}
\tilde{\mathcal{C}}(t)\approx 2n\gamma\int\limits_{-\infty}^{n}\frac{\cos\left[{\frac{\delta\epsilon}{n\hbar}(k-n)\cdot
t}\right]}{1+\alpha^{2}{(k-n)}^2}dk = \tilde{\mathcal{C}}(0)e^{-\frac{t}{\tau_{R}}}
\label{cosint}
\end{eqnarray}
with
\begin{eqnarray}
\label{C0}
\tilde{\mathcal{C}}(0)=\frac{2\pi\lambda^2\Delta
  E^2nN}{{\hbar}^2\alpha},\quad\quad\quad \tau_{R}=\frac{n\hbar\alpha}{\delta\epsilon}
\end{eqnarray}
This is obviously an
exponential decay and thus indicates the applicability of an adequate
BE. To countercheck this result, i.e., the validity of the above
approximations, we compute the time evolution of some current
eigenstate occupation number by solving the time dependent 
SE. The result is shown in the following figure (Fig. \ref{decay}).
In Fig. \ref{decay}  we use for the theoretical curve the results from the
above analysis of $\mathcal{C}(t)$. Obviously there is rather good
agreement between theory and numerics.  Due to the fact that our
theory just predicts only one relaxation time $\tau_{R}$ we conclude that for
our model the ``relaxation time approximation'' seems to be
valid. This finding enables 
as to specify the appropriate matrix of scattering rates for our model: 
\begin{eqnarray}
R_{ij}=-\frac{1}{\tau_{R}}\left(\delta_{ij}-\frac{1}{n\cdot N}\right), 
\end{eqnarray}
where $\delta_{ij}$ denotes Kronecker's delta. After all this analysis
it is justified to state that the microscopic
dynamics of the current eigenmode populations are consistent
with a BE-description as given by (\ref{BEdischom}) with the above matrix of scattering
rates $R_{ij}$.   The equilibrium state $\Phi^0_j$
is specified by $\sum_j R_{ij}\Phi^0_j=0$. Thus one finds
$\Phi^0_j=1/nN$. All other states which are ``orthogonal'' to the
equilibrium state ($\sum_j\Phi_j=0$) correspond to eigenvectors of
$R_{ij}$ with the eigenvalue $-1/\tau_{R}$. Hence the inverse $R_{ij}$
without the null-space is simply given by
$R_{ij}^{-1}=-\tau_{R}$. 
\noindent With those results we may eventually evaluate the transport coefficient
$\kappa$ according to (4). Consequently we insert in (4) for $v_n$ the velocity 
of the quasiparticles cf. (\ref{velo}) for which are approximatively
$v_n\approx \frac{j_{n}}{\Delta E}$. Here we exploit that
$\langle j_n|\hat H|j_n\rangle\approx\Delta E$. Plugging now all
the results into (4) yields:
\begin{eqnarray}
\kappa&=& \frac{\tau_{R}}{n\cdot N\cdot\Delta E^{2}}\sum_{k}j_{k}^{2}\nonumber \\ 
\label{kappa2}
       &=& \frac{\tau_{R}}{n\cdot N\cdot\Delta E^{2}}\text{Tr}\left\{\hat{J}^{2}\right\}, \nonumber\\
\end{eqnarray}
where we exploited the invariance of traces with respect to unitary
transformations. Since Tr$\left\{\hat{J}^{2}\right\}$ is identical with the 
current-autocorrelation function at time $t=0$, i.e., $\tilde{\mathcal{C}}(0)$,
we may use (\ref{C0}) to find
\begin{eqnarray}
\label{diff}
\kappa=\frac{2\pi\lambda^2n}{\hbar\delta\epsilon},
\end{eqnarray}
This is the diffusion coefficient one obtains through an
analysis based on an BE. If it is inserted into (\ref{rateequation}) a simple equation
of motion for the populations of the subunits results. To check
whether the dynamics
\begin{figure}
\centering
\includegraphics[width=80mm]{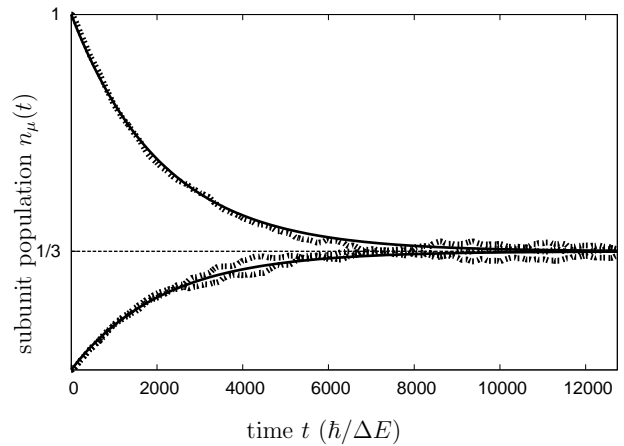}
\vspace*{1mm}
\caption{\label{decay2}  Evolution of the occupation number of the subunits ($n_{\mu}(t)$) 
for a pure initial state featuring $n_{1}(0)=1$ ($n_{2,3}(0)=0$). Dashed lines correspond 
to numerical solutions of the SE, solid lines to theory, cf. (\ref{rateequation}), (\ref{diff}). 
The system parameters are the same like in Fig. \ref{decay}.}
\end{figure}
produced by (10) coincide with the dynamics of the populations of subunits ($n_{\mu}(t)$)
obtained from direct numerical integration of the SE, 
we computed both. The result is displayed in
Fig. \ref{decay2}. Obviously the agreement is rather good. Thus two conclusions
can be drawn: i.) The model indeed shows
diffusive behavior ii.) The diffusive behavior may be interpreted in
terms of scattering quasiparticles. The scattering has to be treated
as proceeding in
such a way that the assumption of molecular chaos or the
Sto\ss{}zahlansatz apply. As a consequence, the macroscopic dynamics may be computed
from a BE. 
\section{Summary and discussion}
\label{sec-5}
We mainly demonstrated that the dynamics of a special class of finite
modular quantum systems are to some extend in accord with the dynamics
generated by an adequately set up BE. The BE here essentially appears
as a rate equation rather than as an evolution equation for the
phase-space density. The  occupation numbers of
current eigenmodes of the quantum system could be shown to obey this
rate equation. Furthermore the
occupation numbers of the local subunits (``modules'') of the quantum
system evolve diffusively, with exactly the same diffusion coefficient
that one gets from analyzing the behavior of long-wavelength
hydrodynamical modes of the BE. 

However, we consider it crucial, that the above described
applicability of a BE to modular quantum systems does not hold in
general. In the example considered in this paper the applicability has
been enforced by the special form of the interaction as described in
(\ref{Cauchy}). This special form contains no restriction regarding
the phases of the transition (interaction) matrix elements but
requires that there weights essentially fall off in a Lorentzian shape
with  energy differences getting larger. A statement about a the
``typicality'' of such interactions can hardly be made, but as a
mathematical condition the special form appears quite restrictive. In
contrary to this large classes of those modular quantum systems exhibit diffusive
behavior with respect to occupation numbers of their subunits even if
the interaction does not feature the above form. This implys that there may be a (large?)
class of systems in general that exhibit diffusive behavior but cannot
be described in terms of a BE picture, i.e., the concept of scattering
quasiparticles may, strictly speaking, be inapplicable. Most
investigations of transport based on the Kubo formula focus on the question whether
the integral over the current-autocorrelation function is finite, not
on whether the correlation function decays exponentially. But the
latter would be needed for the applicability of a BE.

Also the fact that a translationally invariant ``one-particle'' system
exhibits diffusive behavior requires explanation. According to
standard solid-state theory excitations should correspond to free
``lattice-particles'' featuring a dispersion relation depending on the
periodic Hamiltonian. Thus transport is expected to be
ballistic. However, in the case of a rather small amount of subunits
and a large amount of ``orbitals'' per subunit the band structure
looks more like a disconnected set of points in the $E(k)$ vs. $k$ diagram,
rather than the usual set of smooth dispersion relations. In the same
case eigenstates of the current operator do not coincide with Bloch
energy eigenstates. They only do coincide in the limit of the number of
subunits going to infinity in which one then also gets smooth
dispersion relations. So for a comparatively small number of
subsystems the current eigenstates are no stationary states. Thus it appears to be
especially the limit of small numbers of subunits that yields the
diffusive behavior. (We expect that to vanish in the limit of
infinitely many subunits, investigations in that direction are
currently being done.) Thus, for the case of a small number of
subunits we have the seemingly paradox situation that a
translationally invariant one particle system maps onto a BE in which
diffusivity only arises from external scatterers. This, however, one
routinely only expects for systems featuring disorder, defects or
impurities.

\begin{acknowledgments}
We thank K. B\"arwinkel and J. Schnack for
fruitful discussions. Financial support by the Deutsche Forschungsgemeinschaft and the Graduate College 695
``Nonlinearities of optical Materials'' is greatfully acknowledged.
\end{acknowledgments}

\bibliographystyle{}	

\end{document}